\documentclass[12pt]{article}

\usepackage{amsmath}
\usepackage{amsfonts,amssymb,amsthm,overpic}
\makeatletter
\newcommand{\diam}{\mathop{\operator@font diam}}
\usepackage{setspace}
\usepackage{multicol}
\usepackage{multirow}
\usepackage[compact]{titlesec}
\usepackage{xcolor}

\usepackage{epsfig}

\newtheorem{theorem}{Theorem}[section]

\newtheorem{lemma}{Lemma}[section]
\newtheorem{corollary}{Corollary}[section]

\begin{document}

\title{\Huge{\textsc{On null geodesically complete spacetimes uder NEC and NGC; is the Gao-Wald ``time dilation'' a topological effect?}}}

\author{Kyriakos Papadopoulos$^1$\\
\small{1. Department of Mathematics, Kuwait University, PO Box 5969, Safat 13060, Kuwait}\\
\\
E-mail: \textrm{ kyriakos@sci.kuniv.edu.kw}
}

\date{}

\maketitle

\begin{abstract}

We review a theorem of Gao-Wald on a kind of a gravitational ``time delay'' effect in null geodesically complete spacetimes under NEC and NGC, and we observe that it is not valid anymore throughout its statement, as well as a conclusion that there is a class of cosmological models where particle horizons are absent, 
if one substituted the manifold topology with a finer (spacetime-) topology. Since topologies of the Zeeman-G\"obel class incorporate the causal, differential and conformal structure of a spacetime, and there are serious mathematical arguments in favour of such topologies and against the manifold topology, there is a strong evidence that ``time dilation'' theorems of this kind are topological in nature rather than having a particular physical meaning.

\end{abstract}

{\bf AMSC:} 83–XX, 83F05, 85A40, 54–XX    \\
{\bf keywords:} Gao-Wald Gravitational ``Time Delay'', Null Energy Condition, Null Generic Condition, Null Geodesic Completeness, Strong Causality, 
Alexandrov Topology, Closed Diamonds, Compactnes, Zeeman-G\"obel topologies

\section{Preliminaries.}

Gao and Wald have introduced two theorems on gravitational ``time delay'' (see \cite{Gao-Wald}). Here we will only focus on the following one.

\begin{theorem}[Gao-Wald] Let $(M,g_{ab})$ be a null geodesically complete spacetime, satisfying the null energy condition (NEC) and the null generic condition (NGC). Then, given any compact region $K \subset M$, there exists another compact region $K'$ containing $K$, such that if $q,p \notin K'$ and $q \in J^+(p)-I^+(p)$, then any causal curve $\gamma$ connecting $p$ to $q$ cannot intersect the region $K'$.
\end{theorem}

For a detailed treatment of the chronological future and past, respectively $I^+(p)$ and $I^-(p)$, of an event $p$ in a spacetime, we refer to Penrose \cite{Penrose-Topology}, Definition 2.6, p. 12. For the Alexandrov topology on a spacetime,  same reference, Definition 4.22, p. 33; for strong causality Theorem 4.24, p. 34 and for global hyperbolicity Definition 5.24, p. 48. The NEC and NGC are affecting a spacetime as follows:  if a null geodesically complete spacetime satisfies the NEC and NGC, then every null geodesic in the spacetime will contain a pair of conjugate points (for a detailed treatment on conjugate points, NEC, NGC, and the related theorems, we refer to \cite{Hawking-Ellis}).

For the definition of a sliced space we refer to article \cite{Sliced-Spaces}, from where we will make use of the following theorem.

\begin{theorem}\label{1}
 Let $(V,g)$ be a Hausdorff sliced space, where $V =M \times \mathbb{R}$, $M$ is an $n$-dimensional manifold and $g$ is the $n+1$ Lorentz metric in $V$. Then, $(V,g)$ is globally hyperbolic if and only if  $T_P \equiv T_A$ in $V$, where $T_P$  and $T_A$ stand for the product and Alexandrov topologies of $V$, respectively.
\end{theorem}

We will also need the following theorem from \cite{Sliced-Cotsakis}, which gives conditions for global hyperbolicity of a spacetime to be equivalent to null geodesic and timelike geodesic completeness. For the definition of a trivially sliced space we refer to the same article, \cite{Sliced-Cotsakis}.

\begin{theorem}\label{2}
Let $(V,g)$ be a trivially sliced space. Then, the following are equivalent:

\begin{enumerate}

\item The spacetime $(V,g)$ is timelike and null geodesically complete.

\item The spacetime $(V,g)$ is globally hyperbolic.

\end{enumerate}
\end{theorem}

For the definition of the Product topology, one can read, for example, \cite{Dug} while for the definition of the Fine topology, its general relativistic analogue and the Path topology we refer, respectivey, to \cite{Zeeman}, \cite{Gobel} and \cite{Hawking}. In the latter three references, and especially in \cite{Gobel} and \cite{Hawking}, one can read strong arguments against the manifold topology for a spacetime and justify the reason why Zeeman-G\"obel topologies (that is, general relativistic analogues of the topologies that are mentioned in \cite{Zeeman}) are more natural for a spacetime manifold than the manifold topology itself. Throughout the text, we will denote the class of Zeeman-G\"obel topologies by $\mathfrak{Z}-\mathfrak{G}$. So, if $Z \in \mathfrak{Z}-\mathfrak{G}$, then $Z$ will be any spacetime topology definited in \cite{Gobel}.

\section{Some Further Remarks on Gao-Wald's Theorem.}

Combining Theorem \ref{1} and  Theorem \ref{2}, we deduce that:

\begin{corollary}\label{corollary}
 $(V,g)$ is a trivially sliced Hausdorff space where $T_P \equiv T_A$ in $V$, iff $(V,g)$ is timelike and null geodesically complete.
\end{corollary}

The Theorem of Gao-Wald could then be restated as follows:
\begin{theorem}\label{Gaoreviewed}
If $(V,g)$ is a trivially sliced Hausforff space, where $T_P \equiv T_A$ in $V$, and where the NEC and NGC are satisfied, then for any compact set $K \subset V$ there exists a compact set $K'$ containing $K$, such that if $q,p \notin K'$ and $q \in J^+(p)-J^-(q)$, then any causal curve $\gamma$ connecting $p$ to $q$ cannot intersect $K'$.
\end{theorem}

We note that under the statements of Theorem \ref{Gaoreviewed} one could substitute the word ``compact set $K$'' (or $K'$) with a ``closed diamond'' $J^+(a) \cap J^+(b)$, for some arbitary $a,b \in V$, because under global hyperbolicity, closed diamonds are compact (see Penrose, \cite{Penrose-Topology}).This answers partially to the remark of the authors in \cite{Gao-Wald} about the vastness of $K'$; under appropriate conditions, like those in Theorem \ref{Gaoreviewed} one could ``reduce'' the size of $K'$ to a closed diamond $J^+(c) \cap J^-(d)$, for some appropriate $c,d \in V$.

Since $V$ is a product of an $n$-dimensional manifold  with $\mathbb{R}$, it is natural to consider the product topology $T_P$, on $V$. Once again though (just like G\"obel objects against the manifold topology in \cite{Gobel}), one should not ignore that $V$ is also equipped with  $g$, its $n+1$-Lorentz metric, which is an extra structure on $V$. So, even though Gao-Wald gravitational time delay theorem in its original statement or in \ref{Gaoreviewed} is interesting from a geometrical and topological perspective, it is evident that its physical meaning is artificial and it is due to the use of the manifold topology instead of an appropriate spacetime topology. We will comment on this, in more detail, in the section that follows.

\section{The Gao-Wald Gravitational ``Time Delay'' Theorem Fails to hold Under $\mathfrak{Z}-\mathfrak{G}$.}

\begin{lemma}\label{compact}
Let $T_1$ and $T_2$ be two topologies on a set $X$. Let also $T_1$ be finer than $T_2$ (in the sense that $T_1$ has more open sets than $T_2$). Then, the set $K_1$, of all compact sets under the topology $T_1$, will be a subset of the set $K_2$, of all compact sets under $T_2$.
\end{lemma}

For the proof of Lemma \ref{compact}, just consider a compact set $K \in K_1$ and an open cover of $K$ with respect to topology $T_2$. Since $T_2$ is coarser than $T_1$, this fixed cover under $T_2$ wil be a cover under topology  $T_1$ as well. So, there exists a finite subcover in $T_1$ covering $K$ and thus $K$ is compact with respect to $T_2$, too.

Using the result of Lemma \ref{compact}, we see that Theorem 1.1, of Gao-Wald, is not valid if we substitute compactness with respect to the Manifold topology $T_M$ with compactness with respect to any finer $Z$ topology in the class $\mathfrak{Z}-\mathfrak{G}$.

In particular, let us consider our spacetime $M$ under a topology $Z \in \mathfrak{Z}-\mathfrak{G}$. If $K$, in Theorem 1.1, is compact under $T_M$, then it will not be necessarily compact with respect to $Z$, thus $S_k$ will not be compact, too. In the proof of Theorem 1, of  \cite{Gao-Wald}, $S$ consists of points in the tangent bundle of the spacetime where the tangent vector is a future directed null vector, $K$ is a compact set in the spacetime and $S_K$ the restriction of $S$ to points corresponding to $K$. Gao-Wald use $S_k$ to build a compact set $K'$ containing $K$; if $S_K$ is not compact, then the theorem will fail to hold.
If, on the other hand, $K$ is compact with respect to $Z$, then it will necessarily be compact with respect to $T_M$ as well. So, $S_k$ will be compact with respect to $T_M$ but, again, not necessarily compact under $Z$. Thus, if we decide to equip our spacetime with a finer (and more meaningful) spacetime topology, the gravitational time delay effect of Theorem 1.1 will siege to exist.

\begin{theorem}\label{last}
Let $(M,g_{ab})$ be a null geodesically complete spacetime, satisfying the NEC and NGC. If $M$ is equipped with a topology $Z$ in the class $\mathfrak{Z}-\mathfrak{G}$, of Zeeman-G\"obel topologies, then given any compact region $K \subset M$, there does not necessarily exist a compact region $K'$ containing $K$ to fullfil the gravitational time delay effect which states that: if $q,p \notin K'$ and $q \in J^+(p)-I^+(p)$, then any causal curve $\gamma$ connecting $p$ to $q$ cannot intersect the region $K'$.
\end{theorem}

In addition to Theorem 1 in \cite{Gao-Wald},  the authors present a  corollary (Corollary 1, \cite{Gao-Wald}), with the conclusion that there is a class of cosmological models where particle horizons are absent. It is easy to see that Theorem \ref{last} also implies the invalidity of this conclusion under a topology of the class $\mathfrak{Z}-\mathfrak{G}$.

In summary, the Gao-Wald gravitational time delay effect, in the particular case of Theorem 1.1, is due to the choice of the manifold topology $T_M$, against a more natural spacetime topology. The same holds for the establishment of the absense of particle horizons in a class of cosmological models, which depends on Theorem 1.1.  In addition to recent results (e.g. see \cite{Singularities})  which show that the basic theorems on spacetime singularities cannot be formed under topologies in the class  $\mathfrak{Z}-\mathfrak{G}$, due to the failure of the Limit Curve Theorem to hold under such topologies, we now have further evidence against the use of the manifold topology on a spacetime as a ``natural'' topology for a spacetime.

\end{document}